\documentclass[
reprint,
superscriptaddress,
amsmath,
amssymb,
pra,
floatfix,
twocolumn,
]{revtex4-1}
\usepackage{tikz}
\usetikzlibrary{quantikz}
\usepackage{subcaption}
\usepackage{float}
\usepackage{hyperref}
\usepackage{graphicx}
\newcommand{\stab}[1]{\langle #1\rangle}
\usepackage{braket}
\begin{document}
\title{Towards Demonstrating Fault Tolerance in Small Circuits Using Bacon-Shor Codes}
\author{Ariel Shlosberg}
\thanks{A. Shlosberg and A. M. Polloreno contributed equally to this work.}
\author{Anthony M. Polloreno}
\thanks{A. Shlosberg and A. M. Polloreno contributed equally to this work.}
\author{Graeme Smith}
\affiliation{JILA,  University  of  Colorado/NIST,  Boulder,  CO,  80309,  USA}
\affiliation{Department  of  Physics,  University  of  Colorado,  Boulder  CO  80309,  USA}
\date{\today}
\begin{abstract}
\noindent 
Quantum error correction is necessary to perform large-scale quantum computations in the presence of noise and decoherence. As a result, several aspects of quantum error correction have already been explored. These have been primarily studies of quantum memory\cite{chen2021exponential, andersen2020repeated}, an important first step towards quantum computation, where the objective is to increase the lifetime of the encoded quantum information. Additionally, several works have explored the implementation of logical gates\cite{egan2020fault, linke2017fault, 2102.13071}. In this work we study a next step - fault-tolerantly implementing quantum circuits. We choose the $[[4, 1, 2]]$ Bacon-Shor subsystem code, which has a particularly simple error-detection circuit. Through both numerics and site-counting arguments, we compute pseudo-thresholds for the Pauli error rate $p$ in a depolarizing noise model, below which the encoded circuits outperform the unencoded circuits. These pseudo-threshold values are shown to be as high as $p=3\%$ for short circuits, and $p=0.6\%$ for circuits of moderate depth. Additionally, we see that multiple rounds of stabilizer measurements give an improvement over performing a single round at the end. This provides a concrete suggestion for a small-scale fault-tolerant demonstration of a quantum algorithm that could be accessible with existing hardware.
\end{abstract}

\maketitle
\section{Introduction}\label{sec:intro}
In order to perform useful quantum computational tasks that require many qubits or deep circuits, it is necessary to use quantum error-correcting codes that have the ability to handle faulty physical hardware\cite{shor1996fault}. Historically, fidelities of quantum devices have been too far below the necessary thresholds to warrant any attempts to demonstrate fault-tolerant quantum computation. However, slow and steady progress has pushed both single and two-qubit gate infidelities as low as $10^{-3}$\cite{Gaebler2016, Ballance2016, arute2019quantum}. It has therefore become an important theoretical challenge to develop initial small-scale quantum error correction (QEC) experiments to validate the theory. 

The aim of error correction is to make logical qubits and gates that have lower effective error rates than bare physical systems. Over the past decade, there have been multiple non-fault-tolerant QEC code demonstrations using platforms such as superconducting qubits, trapped-ion systems, and microwave modes\cite{Nigg302,Crcoles2015,campagneibarcq2020quantum, fluhmann2019encoding}. While these have all been important building blocks towards large-scale quantum computation, we are more generally interested in fault-tolerant schemes, which effectively mitigate the errors that occur throughout the operation of a QEC scheme. In addition to fault-tolerant quantum circuits, other hardware-efficient fault-tolerant schemes have been proposed, such as performing stabilizer extraction through coupling to nonlinear-oscillators\cite{PhysRevX.9.041009}. In this paper, however, we will restrict our attention to the quantum circuit model. In recent years, experiments have studied the near-threshold performance of quantum error-correcting and error-detecting schemes. This has included exploring multiple rounds of error-detection/correction in small surface codes\cite{andersen2020repeated} and a large repetition code\cite{chen2021exponential}. Recent trapped ion experiments have additionally yielded promising results. For example, Egan et. al implemented circuits designed to be fault-tolerant and studied the implementation of logical gates in a $[[9, 1, 3]]$ Bacon-Shor code\cite{egan2020fault}, although the logical gates were still outperformed by the physical gates. In another experiment, Ryan-Anderson et. al prepared codewords of the [[7, 1, 3]] color code and demonstrated several rounds of error correction, also with circuits designed to be fault-tolerant\cite{2107.07505}. 

These experiments have been primarily quantum memory experiments, which aim to preserve quantum information, rather than process it. However, the ultimate goal of QEC codes is to perform computations on faulty physical hardware in addition to preserving quantum memory. Therefore, we must consider QEC schemes that can handle errors that occur in the logical manipulation of quantum data. 

In this paper, inspired by Gottesman\cite{gottesman2016quantum}, we explore a small quantum error-detecting scheme to demonstrate a limited form of fault-tolerant quantum computation. The advantage of considering quantum error-detection, as opposed to error-correction, is that the physical circuit we take consists of only 5 physical qubits, greatly simplifying the experimental procedure. In particular, the $[[4,2,2]]$ code requires only one additional ancilla to encode two logical qubits. By treating one of those logical qubits as a gauge qubit, we can construct a Bacon-Shor subsystem code, which has fault-tolerant error-detection circuits. These error-detection circuits can be repeated intermittently throughout a computation to detect errors as they occur and to significantly reduce the likelihood of multiple undetectable errors.

\begin{figure*}
  \includegraphics[width=2\columnwidth]{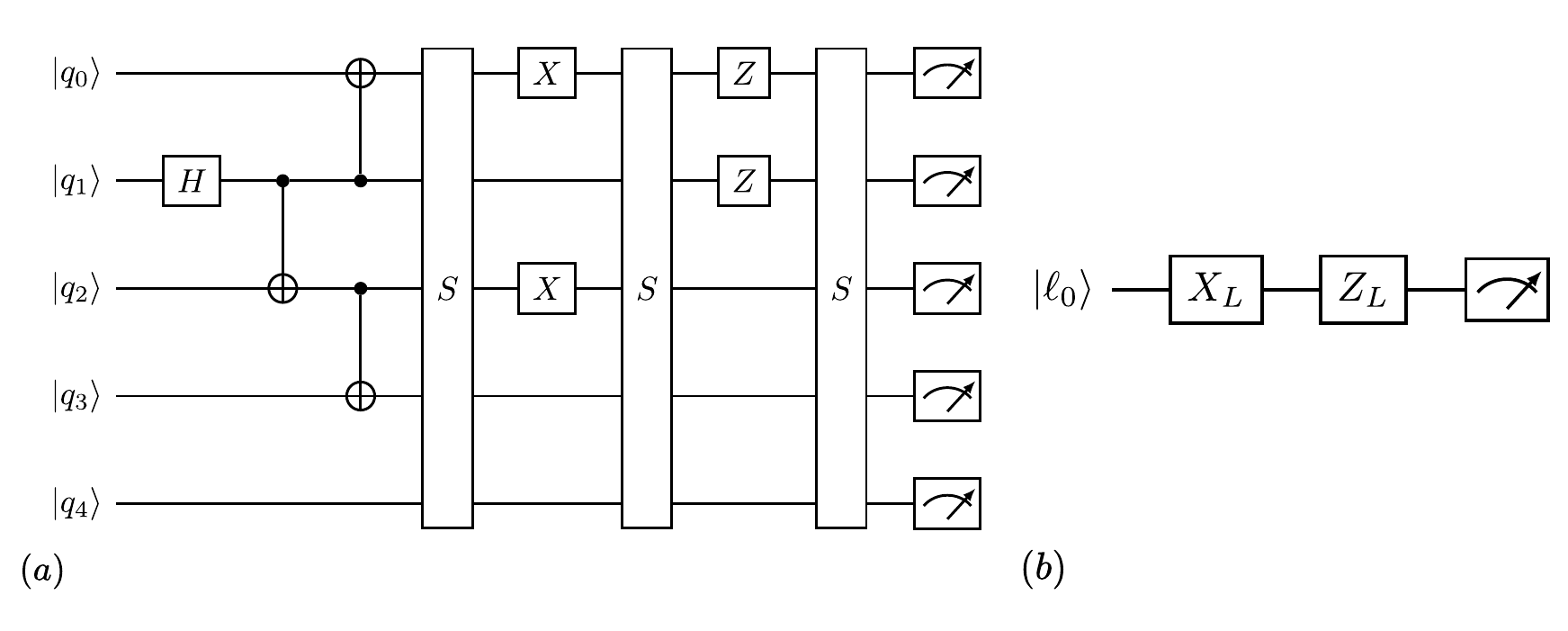}
  
  \caption{Comparison of the encoded fault-tolerant circuit to the bare circuit. a) A physical, fault-tolerant circuit encoding the logical circuit shown in Fig. b. Following state preparation and each logical gate, a syndrome measurement is applied, denoted by $S$. This corresponds to the case of one gate per stabilizer measurement round. Final measurement outcomes are post-selected on obtaining trivial error syndromes. The logical operators $X_L$ and $Z_L$ in Fig. b are implemented via the physical operators $XIXI$ and $ZZII$. b) Qubit $\ell_0$ is acted on by logical gates $X_L$ and $Z_L$, and then measured. We say an experiment demonstrates fault tolerance if a family of logical circuits (one member of which is shown in b)) has a higher success probability on average when implemented with encoded circuits (one example shown in a)), rather than with the corresponding unencoded circuits (a single X followed by a single Z, in this example).}
  \label{fig:fault-tolerance}
\end{figure*}

Gottesman defines fault tolerance in small experiments as being when any encoded circuit, from a given family of circuits $\mathcal{C}$, outperforms its unencoded counterpart\cite{gottesman2016quantum}. That is to say that every encoded circuit in $\mathcal{C}$ must have a lower error rate than the error rate of the equivalent bare circuit. Given the computational difficulty of testing every element of $\mathcal{C}$, which is exponentially large for any given circuit depth, we loosen the notion of fault tolerance to only requiring that the average encoded circuit from a uniform random sample collected from $\mathcal{C}$ have increased performance. For our family we have chosen circuits which consist of $X, Z$ and $H$ gates, that are transversally implementable in the code we consider, and thus fault tolerant. An example of such a circuit is shown in Fig.~\ref{fig:fault-tolerance}, consisting of a single $X$ and $Z$ gate.

In what follows, we use threshold to mean the pseudo-threshold for the particular code being considered. We take two approaches to estimate the fault tolerance threshold for the four-qubit Bacon-Shor code. First, we study depolarizing noise numerically. By simulating circuits of varying depth, we find a threshold for the depolarizing parameter as high as $0.6\%$ for deep circuits. Next, we consider a site-counting argument which uses a stochastic noise model, where errors occur at each site within a circuit with some probability\cite{aliferis2005quantum}. This site-counting argument is more pessimistic than the depolarizing model - it fails to account for benign sets of errors, and assumes that any error that occurs is the worst possible one. Even so, this analysis yields a threshold of $0.2\%$. These analyses show that the threshold for seeing increased performance over the unencoded circuit could be achievable using existing hardware.

\section{Bacon-Shor Subsystem Codes}\label{sec:theory}
The fundamental idea of QEC is to encode the state of a given set of qubits, with Hilbert space $\mathcal{H}_S$, into a larger physical system. Generically this entails mapping states on $\mathcal{H}_S$ into the codespace, $\mathcal{H}_{\text{c}}$, of a higher-dimensional Hilbert space that can be decomposed as $\mathcal{H} = \mathcal{H}_{\text{c}}\oplus\mathcal{H}_{\text{c}_{\perp}}$. Physical errors that move the state of the qubits out of the codespace, and into $\mathcal{H}_{\text{c}_{\perp}}$, can be detected and in some cases corrected. A subsystem code breaks apart the codespace into a product of subsystems($\mathcal{H}_{\text{c}} = \mathcal{H}_L\otimes \mathcal{H}_{G}$)\cite{aliferis2007subsystem, bacon2006operator}. One of the subsystems, $\mathcal{H}_L$, is taken as the logical subsystem and the remaining subsystem, $\mathcal{H}_G$, contains gauge degrees of freedom. In this construction, the state on $\mathcal{H}_G$ does not affect the encoded logical state.

A large class of QEC codes are the stabilizer codes, where logical qubits are encoded as codewords that are in the shared $+1$-eigenspace of a set of commuting Pauli operators\cite{quant-ph/9705052}. The two-dimensional Bacon-Shor code is a subsystem code with a codespace defined by an associated stabilizer group on a square lattice\cite{bacon2006operator}. The stabilizer framework provides a method of detecting whether errors occur in the process of computation, and the division of the codespace into logical and gauge subsystems allows for a simplified scheme of error syndrome extraction. In the case of the Bacon-Shor code, stabilizer measurements can be performed by measuring a set of spatially-local, two-qubit gauge operators, $\mathcal{S}$, whose products contain the stabilizers. The gauge operators do not affect the logical state, but may act non-trivially on the gauge subsystem.

In this work, we consider the four-qubit Bacon-Shor subsystem code, in which there are two elements in the tensor factorization, giving a single logical qubit and a single gauge qubit. We choose the set of gauge operators to be $\mathcal{S} = \stab{IIXX, XXII, ZIZI, IZIZ}$. The logical space of the code is given as the simultaneous +1-eigenspace of the center of $\mathcal{S}$, forming the stabilizer group of the code - in this case $IIII, XXXX, YYYY,$ and $ZZZZ$. The remaining operators will in general act non-trivially on the gauge qubit, meaning that each logical state has in general a family of physical states associated with it.

To see how this works, we will identify the logical operations and states of the code being considered. A logical zero state is given by $\ket{0}_L = \frac{1}{\sqrt{2}}(\ket{0000} + \ket{1111})$. However as we are considering a subsystem code, then there is another orthogonal state that can be associated with logical zero, given by $\ket{\tilde{0}}_L = \frac{1}{\sqrt{2}}(\ket{0011} + \ket{1100})$. Any state in the subspace spanned by $\ket{0}_L$ and $\ket{\tilde{0}}_L$ corresponds to the same logical zero state, and each of these states only differ by the state of the gauge qubit. The associated $[[4,2,2]]$\cite{vaidman1996error, grassl1997codes} code has the same stabilizers as the Bacon-Shor code discussed in this paper. The codewords of the $[[4,2,2]]$ code are
\begin{equation}
\begin{split}
    \ket{00} \rightarrow \frac{1}{\sqrt{2}}(\ket{0000} + \ket{1111})\\
    \ket{01} \rightarrow \frac{1}{\sqrt{2}}(\ket{1100} + \ket{0011})\\
    \ket{10} \rightarrow \frac{1}{\sqrt{2}}(\ket{1010} + \ket{0101})\\
    \ket{11} \rightarrow \frac{1}{\sqrt{2}}(\ket{0110} + \ket{1001}),\\
\end{split}
\end{equation}
and so we see that $\ket{\tilde{0}}_L$ is generated by acting on $\ket{0}_L$ with the gauge operator $XXII \in \mathcal{S}$. The operator $ZIZI$ similarly performs a $Z$ rotation on the second logical qubit in the $[[4,2,2]]$ code (which is taken as the gauge qubit), but does not act on the logical codespace. This is because gauge operators are logical operators on only one of the qubits in the $[[4,2,2]]$ code, and therefore don't affect the logical state of the qubit encoded in the $[[4, 1, 2]]$ Bacon-Shor code. In addition to gauge operators, there are the logical operators $X_L = XIXI$ and $Z_L=ZZII$, which transition the state between distinct logical codewords ($\ket{0}_L$ to $\ket{1}_L$ in the case of $XIXI$).

While the number of logical qubits is fewer than that associated with the $[[4, 2, 2]]$ code, we get several good properties from using a Bacon-Shor code. First, a convenience of working with this code is that the stabilizer operators are all weight-two, which are generally easier to measure fault-tolerantly. As shown in Fig.~\ref{fig:stab_circuits}, a measurement of all the error syndromes only requires eight two-qubit gates. While these gates may require additional SWAPs in a superconducting architecture, trapped-ion architectures may incur no overhead, as they typically have all-to-all connectivity. This suggests that the required threshold for being able to fault-tolerantly measure the stabilizers will be lower for trapped-ion systems.

\begin{figure}[h]
    \centering
    \includegraphics[width=0.5\columnwidth]{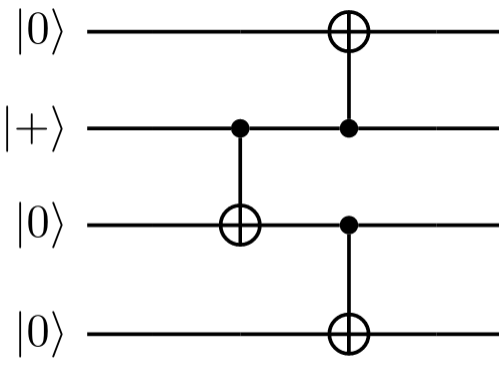}
    \caption{Logical zero ($\ket{0_L}$) preparation circuit for the $[[4, 1, 2]]$ Bacon-Shor code.}
    \label{fig:state_init}
\end{figure}

Finally, the preparation and stabilizer circuits in Fig.~\ref{fig:state_init} and Fig.~\ref{fig:stab_circuits} can be implemented fault-tolerantly. In fact, while the state preparation circuit for the $[4,2,2]$ code requires an ancilla, Fig.~\ref{fig:state_init} shows that the Bacon-Shor code does not, since a single error can at worst propagate to errors on the first and second qubit, or the third and fourth qubit. In both cases, these correspond to performing a gauge operation that changes the state of the gauge qubit, which is irrelevant for this code. The logical codeword $\ket{0_L}$ can thus be prepared through the initialization circuit shown in Fig.~\ref{fig:state_init}.

While a universal gate set is not transversally implementable on the Bacon-Shor subsystem code (meaning that not all single-qubit logical gates can be decomposed entirely into combinations of single-qubit phyical gates), it is possible to transversally implement the logical gates $X_L = X_1X_3$, $Z_L = Z_1Z_2$, and $H_L=\text{SWAP}_{23}H_1H_2H_3H_4$. $\text{SWAP}_{23}$, for our purposes, can be implemented just by updating the labels of the qubits.

\section{Numerics}\label{sec:numerics}

\begin{figure*}
    \centering
    \includegraphics[width=2\columnwidth]{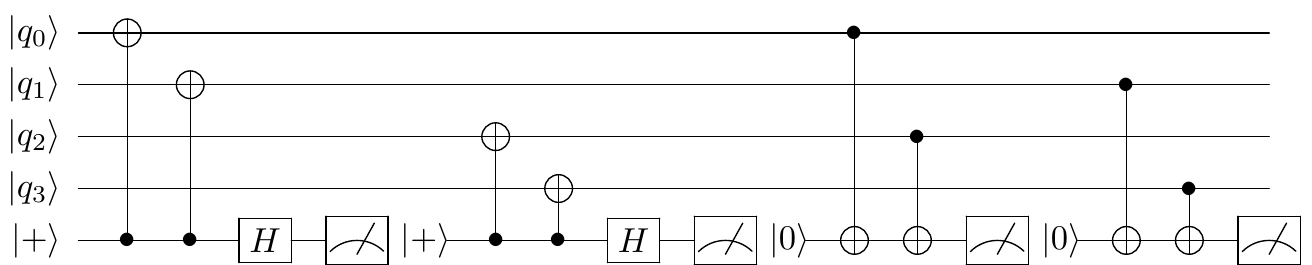}
    \caption{Circuit for extracting error syndromes through gauge measurements ($XXII,IIXX,ZIZI,IZIZ$). Ancilla qubit is reset to $\ket{0}$ and $\ket{+}$ after measurement of each of the gauge operators.}
    \label{fig:stab_circuits}
\end{figure*}

We simulate the four-qubit Bacon-Shor code, described in Section~\ref{sec:theory} with a depolarizing noise model. Throughout each simulation, we repeat rounds of stabilizer measurements, as shown in Fig.~\ref{fig:stab_circuits}, and obtain error syndromes. If $\langle XXXX\rangle = -1$ or $\langle ZZZZ \rangle = -1$, then we detect an error and discard the computation (thereby post-selecting on the trivial error syndromes).

Since this an error-detecting code, we re-run circuits whenever we detect an error. Thus, especially for circuits with many rounds of syndrome measurements, there is a non-trivial chance that a run will be discarded. While the total variation distance to the noiseless output distribution of the logical circuit is a natural metric to compare the encoded and bare circuits, we also divide by the survival probability of the encoded circuit when estimating the threshold. For low enough physical error probabilities, the logical encoding enhances performance despite the use of this conservative threshold estimate.

Finally, at the end of the logical circuit, we measure the physical qubit state destructively in either the computational basis or by performing a logical Hadamard and then measuring in the computational basis. Decoding the observed outcomes on the four physical qubits to a logical state, we compare the fault-tolerant circuit to the bare-qubit computations to get the error-thresholds.

In order to explore fault tolerance, it is necessary to assume all components of the circuit are affected by noise. For the purposes of the simulation, we assume that qubits could be initialized to the $\ket{0}$ state and $\ket{+}$ state noiselessly before logical state initialization. Computational basis state measurements for error syndrome extraction and final logical state determination are also assumed to be noiseless. In theory this assumption can be justified by using extra qubits to prepare a cat state of the circuit output before a noisy measurement. By measuring this cat state and performing a majority-voting scheme, it is possible to suppress the measurement error\cite{preskill1998lecture}. We model the noise after each gate as a depolarizing channel. The single-qubit depolarizing channel has the conventional form
\begin{equation}
    \mathcal{N}_{d}(\rho) = (1-p)\rho + \frac{p}{3}(X\rho X + Y\rho Y + Z\rho Z). 
\end{equation}
We apply this depolarizing channel to the output of each single-qubit gate. Additionally the Bacon-Shor error-detecting code requires using two-qubit entangling gates, CNOTs. At the output of each CNOT, we apply the uncorrelated depolarizing channel to each qubit independently. The error probabilities, defined by parameter $p$, are the same for each depolarizing channel whether applied to the outputs of the one- or two-qubit gates. 

To get a sense of the threshold for this code, we consider logical circuits that are randomly constructed out of the transversal gate set ($X_L,Z_L, H_L$), for varying circuit depths, and acting on the $\ket{0_L}$ state. This corresponds to sampling randomly from the family of circuits mentioned in Sec.~\ref{sec:intro}. We then perform the gauge measurements $\mathcal{S}$ (shown in Fig.~\ref{fig:stab_circuits}) throughout the sequence of applied logical gates.

For each simulated logical circuit depth $d\in\{1,2,5,10,20,30,40,48,60,84,100\}$, we generate 200 random logical circuits per physical depolarizing error parameter. At the output of each circuit, the noisy logical probability distribution $p_L$, which is either $\{p_{L_0},p_{L_1}\}$ or $\{p_{L_+},p_{L_-}\}$, is obtained and the total variation distance, $\delta_L$, from these distributions to the true distribution $p_t$ is calculated as
\begin{equation*}
\delta_L = \frac{1}{2}\sum_{i=1}^2|p_L(i)-p_t(i)|.
\end{equation*}
Similarly, we run a single noisy qubit version of the circuit and compare the output distribution $p_s$ to the true distribution to obtain the total variation distance $\delta_s$. For a fixed gate depth, we plot the average of $\delta_L$ and the average of $\delta_s$ over the $200$ random circuits as a function of the depolarizing error probability. Following \cite{gottesman2016quantum}, we account for the high chances of discarding an encoded circuit run by weighting the quantum error-detecting code variation distance by the post-selection probability, $p_{ps}(p)$, and look for the intersection of $\delta_L/p_{ps}$ with $\delta_s$. We find the intersection points by curve-fitting a quadratic to $\delta_L/p_{ps}$ and a linear fit to $\delta_s$ and then determine the error thresholds.

\begin{figure*}
  \begin{subfigure}[t]{0.45\textwidth}
  \centering
  \includegraphics[scale=.5]{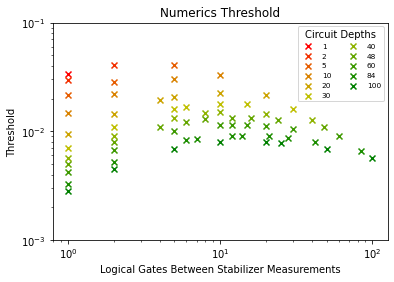}
  \caption{\label{fig:numerics_thresholds}}
  \end{subfigure}
  \begin{subfigure}[t]{0.45\textwidth}
  \centering
  \includegraphics[scale=.5]{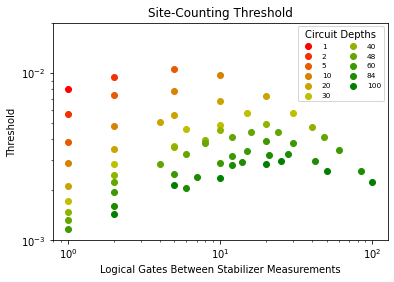}
  \caption{\label{fig:sitecounting}}
  \end{subfigure}
  \begin{subfigure}[t]{0.45\textwidth}
  \centering
  \includegraphics[scale=.5]{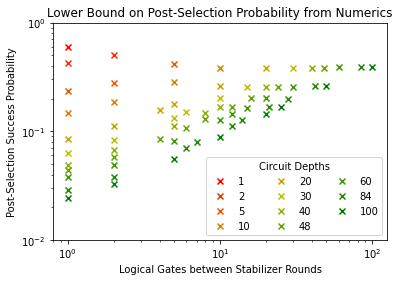}
  \caption{\label{fig:postselect}}
  \end{subfigure}
  \begin{subfigure}[t]{0.45\textwidth}
  \centering
  \includegraphics[scale=.5]{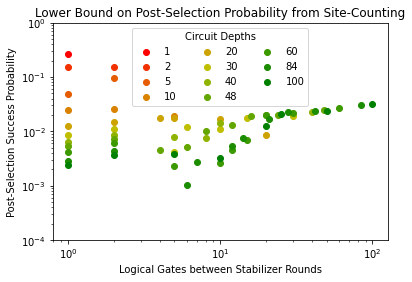}
  \caption{\label{fig:sitecounting_postselection}}
  \end{subfigure}
  \caption{(a) Error probability thresholds at different depths (1,2,5,10,20,30,40,48,60,84,100) with a variety of repeated stabilizer rounds computed through numerics. (b) Error probability thresholds computed through site-counting. While the typical value is an order of magnitude lower than in the case of the numerics, the shape of the bound is qualitatively similar, showing an improved threshold with a nontrivial rate of intermediate measurements. (c) Post-selection probabilities computed through numerics, evaluated at the threshold value.  (d) Post-selection probabilities computed through site-counting, evaluated at the threshold. While the lower bound is more conservative than the one in (c), we again see that the results are qualitatively similar, approaching a fixed value when only one measurement is performed at the end.}
\end{figure*}

As can be seen in Fig.~\ref{fig:numerics_thresholds}, there is an ideal frequency of performing stabilizer measurement rounds that corresponds neither to making a single stabilizer measurement at the end, nor performing error-detection after each logical gate. This is the result of two different phenomena. The first is that the stabilizer circuit is itself noisy and thereby increases the likelihood of generating single-qubit errors in the physical qubits. The second is that there is an increase in the post-selection probability as the error-detection round frequency is decreased, this is shown in Fig.~\ref{fig:postselect}. This is because the circuit is more likely to allow logical errors, but is less likely to find non-trivial error syndromes. The numerics indicate that given the depolarizing error model with the $[[4,1,2]]$ Bacon-Shor code, the physical error threshold is maximized at approximately one round of stabilizer measurement per 15 logical gates. 

\section{Site-Counting}\label{sec:site-counting}
For another estimate of the threshold of this code, we will follow the site-counting argument in \cite{aliferis2005quantum, gottesman2016quantum} closely. We will derive a bound on the logical error rate for an analytic prediction of when we can expect an encoded algorithm to outperform its unencoded counterpart. As discussed in both Sec.~\ref{sec:intro} and Sec.~\ref{sec:numerics}, we consider circuits consisting uniformly of $X_L$, $Z_L$ and $H_L$ gates. These require two, two and four physical gates (as noted previously, SWAP gates can be implemented by relabeling qubits) respectively, so a circuit with $T$ logical gates, will on average contain $(2 + 2 + 4)T/3 = 8T/3$ physical sites where an error can occur. 

We assume perfect initialization of both the $\ket{0}$ and $\ket{+}$ states, perfect measurements and no idle noise. While in principle these will contribute more infidelity to the logical circuit, the contribution will be small. Using the initialization circuit in Fig.~\ref{fig:state_init} with 3 CNOTs, we get $N_A=6$ locations in the logical state preparation circuit where errors can occur. 

The stabilizer measurements require measuring $\stab{IIXX}$ and $\stab{XXII}$, which both require 2 CNOTs and 1 Hadamard, contributing 10 locations for error, and measuring $\stab{IIZZ}$ and $\stab{ZZII}$ requires 2 CNOTs, giving 8 locations for errors. Thus we have $N_B=18$ locations per stabilizer measurement. A logical error only arises if more than one error occurs between measurements. Using site-counting we find that the probability for a logical error, $p_l$, in a circuit with $M$ stabilizer measurements, and $N = \frac{8}{3}T/(M+1)$ physical gates between measurements is bounded as
\begin{multline}
    1 - p_l \geq (1-{N_A + N \choose 2}p^2)(1 - {N_B + N \choose 2}p^2)^{M},
\end{multline}
where $p$ is the physical error rate. Additionally, we can bound the post-selection probability. For each block of the circuit between stabilizers, a bound on the post-selection probability is given by $1 - Sp$ where $S$ is the number of sites per block, giving
\begin{equation}\label{eq:sitecounting-ps}
    p_{ps} \geq (1 - (N_A + N)p)(1 - (N_B + N)p)^M.
\end{equation}
For the unencoded circuit, we have that $p_u\approx Tp$. Finally, to compare the two, we will consider the conditional probability of success. The resulting threshold for when the encoded circuit begins to outperform the unencoded circuit is shown in Fig.~\ref{fig:sitecounting}. We can additionally use site-counting to estimate the rejection probabilities as shown in Fig.~\ref{fig:sitecounting_postselection}. In both cases, the result is qualitatively similar - by repeatedly measuring stabilizers, the post-selection probability decreases exponentially, however by measuring the stabilizers fewer times the post-selection probability increases, due to the potential for errors to collude and become undetectable. In both cases, site-counting gives a more conservative threshold than the numerics. This is due to the looseness of the union bound and because our argument fails to account for benign errors. We could improve our bound by considering benign errors, as in \cite{aliferis2005quantum}.

As in Sec.~\ref{sec:numerics}, we can estimate the optimal number of stabilizer measurements to perform for each circuit depth. Taking the maximum of each curve in Fig.~\ref{fig:sitecounting} we get (1, 2, 5, 10, 20, 15, 20, 16, 20, 28, 25) gates between stabilizer measurements for depths (1, 2, 5, 10, 20, 30, 40, 48, 60, 84, 100) respectively. The site-counting argument is overly pessimistic about the errors contributed by the stabilizer measurements, and so is more sparing in their use, which can be seen by comparing to the 15 logical gates between stabilizer measurements suggested by the numerics in Sec.~\ref{sec:numerics}.

\section{Conclusion}
Quantum error correction is a necessary step towards the development of large-scale quantum computers. Quantum fault tolerance is the demonstration of a quantum error correction scheme that is robust to errors in the error correction schemes themselves. Experimental platforms have recently achieved impressive fidelities that put us near the range that we should consider small quantum codes for both quantum memories and small demonstrations of fault tolerance. Indeed, we have yet to see an unambiguous demonstration of improvement from fault tolerance. In this paper, we have proposed a simple demonstration of fault tolerance that can be used to perform encoded circuits. 

Through both numerics and a site-counting argument, we have shown that Pauli error rates for short circuits (with a depth of five) as high as $p=3\%$ are capable of demonstrating an improvement of the encoded circuit over the unencoded circuit. Moreover, for deeper circuits (with a depth of 100), we find that performing more than one round of gauge measurements (one round per 15 logical gates) provides a threshold error rate as high as $p=0.6\%$, making experimental observation of the threshold potentially realizable on superconducting and trapped ion platforms.

In our analysis, we have assumed that single- and two-qubit gates have similar error rates, modeling two-qubit gates as being subjected to two single-qubit errors. However, state-of-the-art quantum computing platforms have single-qubit error rates that are often smaller by at least an order of magnitude. Nevertheless, the code considered in this paper encodes a single logical qubit. Thus the natural quantity to compare against in a discussion of fault tolerance, and the one we chose to compare against, is another single qubit physical error rate.

There are at least two natural extensions to this work. The first is considering two logical qubits, each encoded in their own four-qubit Bacon-Shor code. In this case, the natural quantity to compare against is the two-qubit gate infidelity, and in the analysis of the logical circuit one could consider single-qubit gates as free. Then one would only count the two-qubit gates used in the state initialization circuit, stabilizer measurements, and to perform entangling operations between the two encoded qubits. This model makes the more realistic assumption of different error rates for gates involving different numbers of qubits. This assumption is particularly relevant in trapped ion systems where single-qubit gate infidelities are exquisitely small\cite{Gaebler2016, Ballance2016}. 

The other straightforward extension of this work is to consider the $[[9,1,3]]$ Bacon-Shor code. This code also encodes a single logical qubit, but is distance three and therefore can correct one error. Thus, rather than simply rejecting circuits as discussed in this paper, we can consider implementing error-correcting circuits which should improve the threshold for deeper circuits.

While this work focused on an uncorrelated depolarizing noise model, an additional direction of research is to consider more realistic noise models for trapped-ion and superconducting architectures. However, we are finally reaching an exciting point in hardware where quantum codes can be implemented and have the ability to outperform bare physical qubit schemes, and are therefore on the brink of scalability. 

\section{Acknowledgements}
This material is based upon work supported by the U.S. Department of Energy, Office of Science, National Quantum Information Science Research Centers, Quantum Systems Accelerator (QSA). AMP is funded under NSF grant number 1734006.
\bibliography{refs}
\end{document}